\documentclass{aastex62}
\usepackage{epsfig,amsmath,amsfonts,amssymb,graphicx,subfigure,enumitem,color}
\usepackage{amssymb}
\usepackage{amsmath}
\usepackage[varg]{txfonts}
\usepackage{natbib}
\usepackage{url}             
\usepackage{graphicx}
\usepackage{float}
\newcommand{\beq}{\begin{equation}}
\newcommand{\eeq}{\end{equation}}
\newcommand{\beqa}{\begin{eqnarray}}
\newcommand{\eeqa}{\end{eqnarray}}
\newcommand{\beqann}{\begin{eqnarray*}}
\newcommand{\eeqann}{\end{eqnarray*}}
\bibpunct{(}{)}{;}{a}{}{,} 

\usepackage{textcomp}
\newcommand{\foo}{\rlap{+}{\texttimes}}
\shorttitle{Forced reconnection by an eruptive prominence}
\shortauthors{Srivastava et al.}

\begin{document}
\title{The prominence driven forced reconnection in the solar corona and associated plasma dynamics}
\author{A.K.~Srivastava}
\affil{Department of Physics, Indian Institute of Technology (BHU), Varanasi-221005, India}
\author{Sudheer~K.~Mishra}, 
\affil{Department of Physics, Indian Institute of Technology (BHU), Varanasi-221005, India}
\affil{Indian Institute of Astrophysics, Sarjapur Main Road, 2nd Block, Koramangala, Bengaluru-560034, India}
\author{P.~Jel\'inek}
\affil{University of South Bohemia, Faculty of Science, Department of Physics, Brani\v sovsk\'a 1760, CZ -- 370 05 \v{C}esk\'e Bud\v{e}jovice, Czech Republic.}

\bigskip
\begin{abstract}
Using the multi-temperature observations from SDO/AIA on 30th December 2019, we provide a signature of prominence driven forced magnetic reconnection in the corona and associated plasma dynamics during 09:20 UT to 10:38 UT. A hot prominence segment erupts with a speed of $\approx$21 km s$^{-1}$ and destabilises the entire prominence system. Thereafter, it rose upward in the north during 09:28 UT to 09:48 UT with a speed of 24 km s$^{-1}$ . The eruptive prominence stretches overlying field lines upward with the speed of 27-28 km s$^{-1}$, which further undergo into the forced reconnection. The coronal plasma also flows in southward direction with the speed of 7 km s$^{-1}$, and both these inflows trigger the reconnection at $\approx$09:48 UT. Thereafter, the east and westward magnetic channels are developed and separated. The east-west reorganization of the magnetic fields starts creating bi-directional plasma outflows towards the limb with their respective speed of 28 km s$^{-1}$ and 37 km s$^{-1}$. Their upper ends are diffused in the overlying corona, transporting another set of upflows with the speed of $\approx$22 km s$^{-1}$ and 19 km s$^{-1}$. The multi-temperature plasma (T$_{e}$=6.0-7.2) evolves and elongated upto a length of $\approx$10$^{5}$ km on the reorganized fields. The hot plasma and remaining prominence threads move from reconnection region towards another segment of prominence in the eastward direction. The prominence-prominence/loop interaction and associated reconnection  generate  jet-like eruptions with the speed of 178-183 km s$^{-1}$. After the formation of jet, the overlying magnetic channel is disappeared in the corona.
\end{abstract}


\section{Introduction}

The Sun$'$s hot and magnetized corona acts as a reservoir for the onset of solar eruptions, spiraling solar wind stream, and their interaction with the planetary magnetic fields constituting its space weather. The one of the fundamental physical processes of the magnetic field in the solar corona, which causes the localized heating, leads eruptions, and provides the kinetic energy for the propulsion of the plasma, is know as magnetic reconnection. On the first instance, the magnetic reconnection is defined as the self-reorganization and relaxation of the complex and twisted magnetic fields in the solar corona leading to the liberation of stored magnetic energy in the constituent plasma due to direct dissipation of the electric current \citep[e.g.,][and references cited there]{2007mare.book.....P,2010RvMP...82..603Y,2014masu.book.....P}. In the solar corona, the magnetic reconnection is one of the major candidates to heat it locally, and to trigger various types of eruptive phenomena and localized plasma dynamics \citep[e.g.,][and references cited there]{1988ApJ...330..474P, 2004ApJ...605..911C,2011LRSP....8....6S,2014ApJ...795..172J,2015RSPTA.37340256K,2016NatCo...711837X}. It is also responsible for the different types of small-scale activities and physical processes in the solar atmosphere such as small-scale Ellerman bombs \citep[e.g.,][and references cited there]{2013ApJ...779..125N, 2014Sci...346C.315P, 2016ApJ...824...96T}, small to large-scale solar jets \citep[e.g.,][and references cited there]{1995Natur.375...42Y, 1997Natur.386..811I, 2015Natur.523..437S,2018NatAs...2..951S}, UV/EUV transients \citep[e.g.,][]{2014Sci...346A.315T}, solar flares \citep[e.g.,][and references cited there]{1994Natur.371..495M,2011LRSP....8....6S,2013NatPh...9..489S,2018ApJ...858...70F}, large-scale filament/prominence eruptions \citep[e.g.,][]{2016NatPh..12..847L, 2016NatCo...711837X}, coronal mass ejections \citep[e.g.,][]{2010ApJ...722..329S}, etc. The magnetic reconnection may also be driven by the external eruptions, flux emergence, etc. \citet{2014ApJ...788...85V} have observed that magnetic reconnection can be driven by the expansion of the coronal mass ejection, and it reconnects with the nearby active region. The erupting fluxrope undergoes the interchange magnetic reconnection with the oppositely directed magnetic field and it is responsible for the eruption. It is similar to the breakout model of the coronal \citep[e.g.,][]{1999ApJ...510..485A, 2010JGRA..11510104C}. The reconnection favoured emerging flux may also be responsible for the triggering of the filament eruption and the associated CME \citep[e.g.,][]{1995JGR...100.3355F, 2000ApJ...545..524C, 2009ApJ...697..913O}. The magnetic reconnection has wide physical implications in the Sun's atmosphere, planetary magnetosphere, heliosphere, and in many other astrophysical objects, e.g., active galactic nuclei, pulsars etc., as well as laboratory plasma \citep{2020JGRA..12525935H}. However, there are several outstanding issues associated with the magnetic reconnection, which are still under debate despite remarkable progress on their study since last three decades both in the frame-work of theory and observations. Few scientific issues are {\it viz.,} determining the geometry of the reconnection region, properties and dynamics of the current sheet, magnetic field configuration and formation of X-point, exact estimation of the reconnection rate, physical process and role of resistive instabilities in the reconnection region, and quantification of resistivity/magnetic diffusivity, etc \citep[e.g.,][and references cited there]{2007mare.book.....P, 2010RvMP...82..603Y,2020SoPh..295..167M,2020JGRA..12525935H,2021SSRv..217...39P}.

The magnetic reconnection also invokes breaking and reconfiguration of the oppositely directed magnetic field lines in the resistive plasma in which they collapse on the X-point and associated current sheet in the localized solar atmosphere \citep{1986JGR....91.5579P, 2005GeoRL..32.6105B}. In the spontaneous magnetic reconnection, current sheet dynamics may be associated with the MHD instabilities e.g., resistive tearing mode instability \citep{1995ApJ...451L..83S, 2001EP&S...53..473S, 2017JPlPh..83e2001V}. The another stable magnetostatic configuration of the current sheet could also be developed in the large-scale solar corona where some external perturbations may trigger the forced magnetic reconnection \citep{2019ApJ...887..137S}. Using the space-borne data from Atmospheric Imaging Assembly (AIA) onboard the Solar Dynamics Observatory (SDO), \citet{2019ApJ...887..137S} have firstly found an observational evidence of the forced reconnection at a considerably high rate, which occurred locally in the solar corona. The observed forced reconnection was generated in the large-scale corona when two oppositely directed magnetic field lines forming an X-point and associated current sheet, are perturbed by an external disturbance generated by the motion of a cool solar prominence. This type of reconnection has only been reported in theory \citep{2005PhPl...12a2904J}, and has never been directly observed in the Sun$'$s large-scale corona. Before this, \citet{2010ApJ...712L.111J} have only observed a microflare activity driven by the forced magnetic reconnection, which they termed as its indirect signature in the solar chromosphere. Recently, \citet{2020A&A...643A.140M} have reported the evolution of sausage waves in the magnetic structures and their role in enabling the forced reconnection in  vicinity in the solar corona.

As seen in the above example, in general, the current sheet may undergo in the process of forced reconnection by some external perturbations generated due to the oscillatory processes, evolution of pulses, coalescence and tearing mode instabilities, etc \citep[e.g.,][and references cited there]{1998PhPl....5.1506V, 2005PhPl...12a2904J, 2017JPlPh..83e2001V, 2019A&A...623A..15P}. There is a significant development in the theory of the forced magnetic reconnection in a variety of magnetized plasma configurations since last three decades \citep[e.g.,][and references cited there]{1985PhFl...28.2412H, 1998PhPl....5.1506V, 2001PhPl....8..132B, 2005GeoRL..32.6105B, 2005PhPl...12a2904J, 2017PhPl...24e2508B, 2019A&A...623A..15P, 2019ApJ...887..137S}. Typically for the solar corona, the forced reconnection is studied by \citet{2005PhPl...12a2904J} as one of the primary candidates for its heating. They have simulated the forced reconnection when a sheared force-free field is perturbed by the slow pulse like disturbances that generate a series of heating events similar to the nanoflare heating. \citet{1998PhPl....5.1506V} have studied earlier the forced reconnection in a force-free magnetic field in a current sheet due to a tearing mode instability, which could mimic the physical scenario of the solar coronal heating. \citet{2019A&A...623A..15P} have modeled a force-free current sheet in the solar corona which allows multiple magnetic islands to be formed and coalesce in order to release the energy rapidly. Recently, using the data-driven MHD modelling, \citet{2019ApJ...887..137S} also showed that even without much development of the typical physical conditions for the reconnection, a magnetic explosion may forcibly occur in a current sheet rapidly due to an external velocity perturbation in order to liberate energy and to heat the localized solar corona. There are simple manifestations in many previous reports that some perturbations may lead the driven magnetic reconnection \citep{2005GeoRL..32.6105B}, however, the forced reconnection possess some specific physical properties that were not observed earlier. The plasma heating in the forced reconnection may be provided by the ongoing external driving and internal reconnection both provided the time-scale of the driver and reconnection matching with each other \citep{2005PhPl...12a2904J}. In the forced reconnection, even if the likely conditions are not present in the localized corona, though the inflows may be driven by the external perturbations, and there will be obvious time-lag between plasma inflows and outflows \citep{2019ApJ...887..137S}. The current sheet can appear obviously in an MHD stable magnetic configuration in the localized corona in response to the some external perturbation, and this further leads to the forced magnetic reconnection \citep{2017JPlPh..83e2001V}. The lesser resistivity may also trigger the reconnection over the stable current sheet under the influence of the external perturbations \citep{2017JPlPh..83e2001V,2019ApJ...887..137S}. Recent observations clearly demonstrated some of these specific physical properties of the forced reconnection, although there is a requirement of further studies on the observations and modeling of this physical phenomenon \citep{2019A&A...623A..15P,2020A&A...643A.140M,2021SSRv..217...38K}.
  
In the present paper, using multi-wavelength observations of the solar corona from the Atmospheric Imaging Assembly (AIA) onboard the Solar Dynamics Observatory (SDO) on 30th December 2019, we observe a cool prominence driven forced reconnection in a dynamical X-point in the off-limb solar corona, and thereafter its responses in form of the hot plasma flows and formation of coronal jet-like dynamics in its surrounding. The dynamical nature of the prominence is governed by eruption of a localized hot segment of the  prominence initially, which further perturbs the overlying magnetic field configuration. Overall, this plasma dynamics triggers the expansion of the cool prominence system that further stretches the overlying coronal magnetic field lines and pushes them into the reconnection region. This dynamical process stretches and expands the overlying coronal field lines, and causes the onset of inflows and subsequently the forced magnetic reconnection in the overlying solar corona. In Sect.~2, we present the observational data and its analyses. The observational results are depicted in Sect.~3. The discussion and conclusions are outlined in the last section.

\section{Observational Data and Analyses}
We analyze the multi-temperature temporal image data from the Atmospheric Imaging Assembly (AIA) onboard the Solar Dynamics Observatory (SDO) as observed on 30th December 2019. The Atmospheric Imaging Assembly (AIA) is onboard SDO \citep{2012SoPh..275...17L}. AIA consists of seven extreme ultraviolet (94, 131, 171, 193, 211, 304, 335 {\AA}), two ultraviolet (1600, 1700 {\AA}) and one visible (4500{\AA}) full disc imager with 1.5$"$ spatial resolution per two pixels. The pixel size of the image data is 0.6$"$, while its cadence is 12 s. We use 94 {\AA}, 131 {\AA}, 171 {\AA}, 193 {\AA}, 211 {\AA}, 304 {\AA}, and 335 {\AA} temporal image data of AIA in the present analysis. We have selected $\approx$2 hour multi-filter/temperature time-series data starting from 09:20 UT on 30$^{th}$ December 2019 to observe a prominence system at the north-eastern limb, and overlying complex loop system. We choose a particular area of 400$"$ by 350$"$ field-of-view (Fig.~1a) ranging from -650$"$ to -1050$"$ in the East-West direction and 400$"$ to 750$"$ in the North-South direction. The basic calibration and normalization of the data were performed by using Solarsoft IDL routine "aia\_prep.pro". \\

We use five channels of SDO/AIA observations (304, 171, 211, 193, 131  {\AA}) to analyze the dynamics of prominence and overlying coronal structures (Fig.~1). The transition region (TR)/upper chromospheric emission at 304 {\AA} is dominated by the He II lines formed between (5-8)$\times 10^{4}$ K, and show the existence of cool plasma. The inner coronal emission at 171 {\AA} is formed around (6-8)$\times 10^{5}$ K to exhibit the appearance of inner coronal plasma. In the present study, the prominence dynamics and associated flows are observed in 304 {\AA}. The forced reconnection region, formation of associated X-point and dynamical current sheet, and overlying coronal magnetoplasma system are best seen in the 171 {\AA} emissions. The coronal magnetic field lines and their dynamics
are collectively visible in 211{\AA}, 193 {\AA}, and 131 {\AA}, which are the high temperature filters of SDO/AIA. The composite images are constructed by combining AIA 304 {\AA} and 171 {\AA} (Fig.~1 right column), and AIA 131 {\AA}, 193 {\AA}, 211 {\AA} (Fig.~1 left column) to observe the behavior of cooler prominence plasma, dynamics of coronal plasma, and the overall dynamics of the forced reconnection region and its surroundings. The AIA 193 {\AA} images are also used to show initially the evolution of the eruption of some hot components of the prominence plasma system
(cf., Fig.~2) that further triggers the expansion of the cool prominence system to force the reconnection in the overlying large-scale solar corona. Figs.~1-5 depict all such analyses and related results, which will be described in detail in the upcoming Sect.~3. The schemtic is also displayed in Fig.~3 to describe the overall dynamical scenario associated with the forced reconnection event driven by the expanding cool prominence system, and stretching of the overlying coronal field lines that further undergo into the reconnection region.

In order to understand thermal structures of the forced reconnection region and associated plasma dynamics, we obtain the Differential Emission Measure (DEM). We map the DEM by using six different temperature AIA filters, i.e, 131 {\AA}, 171 {\AA}, 193 {\AA}, 211 {\AA}, 94 {\AA}, and 335 {\AA} as observed by SDO. We do not use cool and optically thick 304 {\AA} filter to examine  only the relative contributions of the high temperature plasma on the total emission as captured by different hot AIA filters. We have used the method of \citet{2012A&A...539A.146H} to measure the differential emission from the heated prominence material, and also demonstrate its dynamics, surrounding magnetic field, and the reconnection region. The present estimation is based on an automated method, which gives a regularized DEM as a function of temperature (T). In order to obtain the inversion, we implement zeroth-order regularization in the temperature range of log T(K)=5.0 to log T(K)=7.5 over the bin of twenty six temperatures with $\Delta$log T(K)=0.1 intervals. For the selected six hot AIA filters, we compute DEM for the region of interest (ROI) in the selected temperature bins, which is displayed in Fig.~4. 
The dynamics of the forced reconnection region and associated hot plasma outflows, as well as formation of the hot coronal jet-like structure in the vicinity, are observed in the temperature range of log T(K)=6.0-7.2. The detailed results related to the DEM analyses, and their physical implications are described in Fig.~6 and Sect.~3.

\begin{figure*}
\begin{center}
\includegraphics[scale=0.6,angle=0,width=18.5cm,height=19.0cm,keepaspectratio]{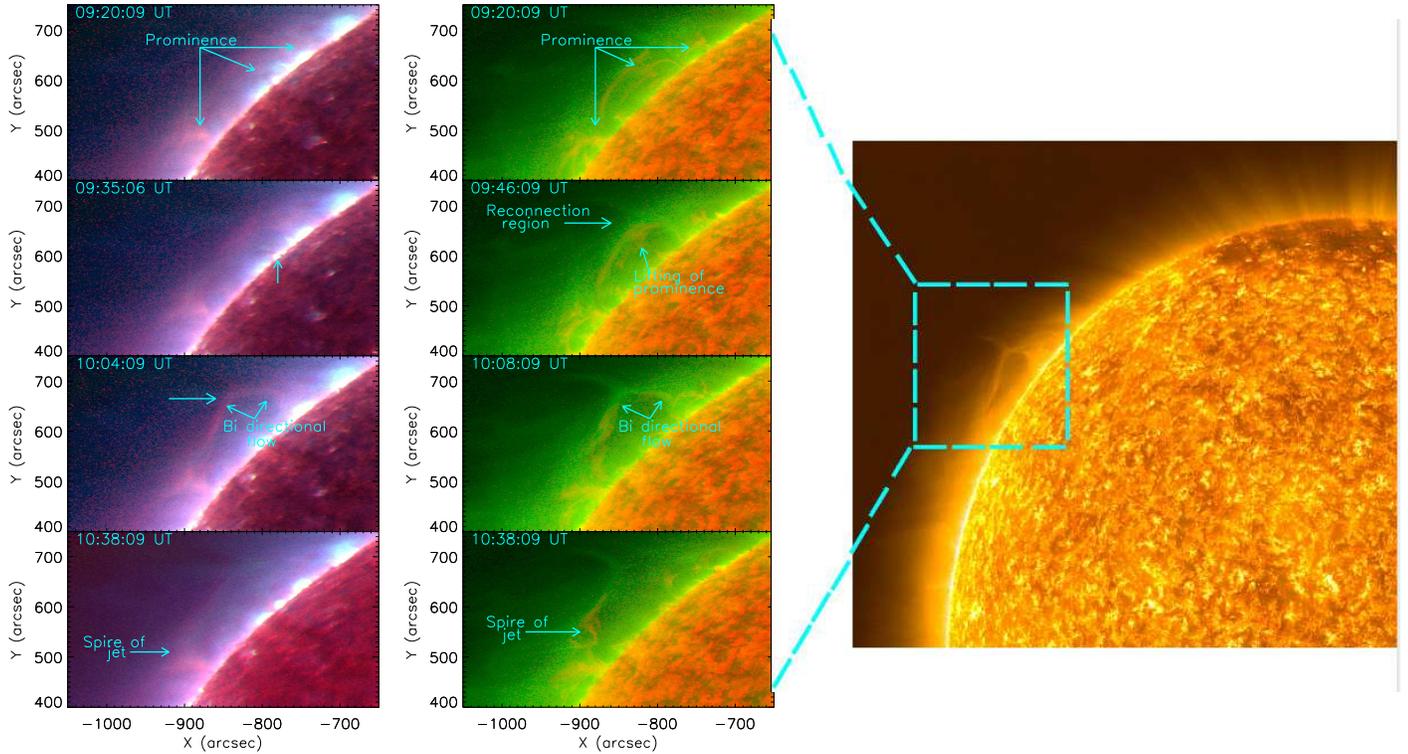}
\end{center}
\caption{Right-panel: A context image of SDO/AIA as a composite image of 304 and 171 {\AA}, shows the forced reconnection region in the cyan dashed-line box. 
Left-panel: The first vertical column of the images consists of the composites of AIA 304 and 171 {\AA}, which show the prominence and overlying coronal magnetoplasma system where the forced reconnection occurred. The second vertical column of the images consists of the composites of AIA 131, 211, 193 {\AA}, which display the dynamics of the stretched overlying coronal magnetic field lines initially responsible for the triggering of the forced reconnection. The animation1a.mp4 (131+193+211) \AA~and animation1b.mp4 (171+304) \AA~of the SDO/AIA composite images show the complete evolution and expansion of the prominence system, eruption of the associated hot segment of the prominence, stretching of the overlying coronal magnetic field configuration, the forced reconnection region, and associated plasma dynamics. It runs from 09:20 to 11:00 UT.} 
\end{figure*}
\section{Observational Results}

The dynamics of the large-scale magnetic field and frozen-in plasma is seen in the North-East part of the off-limb corona during 09:20 UT to 10:38 UT on 30th December 2019 (cf., Fig.~1, right most panel). The off-limb region consists of a cool prominence system (cf., Fig.~1, right-column, 09:20:09 UT). The small localized prominence chunk (reddish-brown) at right-most part is confined below an overlying diffused coronal loop system (green). Some set of large-scale coronal field lines are also present at the right-most of this region, however, their upper end is diffused in the overlying corona. The prominence footpoints are appeared to be active and bright in the higher temperature emissions also (cf., Fig.~1, left-column, 09:20:09 UT). This qualitatively suggests that some localized flux emergence and related heating is enveloping the prominence footpoints, and later makes some parts of them eruptive. We, therefore, observe that the hot component of the plasma present at the active footpoints/pillars of a prominence, is mostly visible at a coronal temperature in AIA 193 {\AA} and AIA 211{\AA} (Fig.~2 and Fig.~1). This plasma structure is started to lift up at $\approx$09:24 UT and causes the overlying prominence associated magnetic field to expand upward. The eruptive prominence further stretches the overlying coronal magnetic field lines that further undergo into the forced reconnection (cf., 10:04:09 UT image in left column of Fig.~1). 
We have used the running difference image of AIA 193 {\AA} to observe the dynamical behaviour of the hot plasma segment of the prominence system (its one low-lying active footpoint) and its possible kinematical trajectory (cf., Fig.~2, top right panel). Later, it is trapped and faded away within the overlying magnetic domain. 
Fig.~2 shows the height-time diagram along the path P1 (top-right panel). It elucidates the eruption of the heated plasma segment of the prominence system and overlying coronal magnetic fields.
The eruption of the prominence's heated plasma segment 
follows a parabolic trajectory, which indicates a failed eruption of it. 
Some internal localized and small-scale reconnections may be responsible for its eruption
, while it is faded and trapped at the later stage due to the overlying magnetic fields. 
This plasma structure moves upward initially by the speed of 21 km s$^{-1}$ (cf., top-right panel of Fig.~2), and further destablises and expands the entire prominence system. The eruptive prominence further stretches the overlying coronal magnetic field lines. These stretched overlying magnetic field lines embedded in the hot plasma move up just above the prominence with a speed of 27-28 km s$^{-1}$ (cf., top-right and bottom panels of Fig.~2). These field lines are also visible in the higher temperature filter of AIA 131 {\AA} (10 MK; cf., left-middle panels of Fig.~2). A path 'P2' has been taken along the expansion of the overlying coronal field lines. We found that the hot plasma embedded in these stretched overlying coronal magnetic fields accelerates toward the X-point with a velocity of  27-28 km s$^{-1}$ (Fig.~2, lower right panel). It is comparable with the scenario observed in AIA 193 \AA~ filter. The cool prominence plasma just moves behind it upward towards the reconnection region.
It is seen that some other set of large-scale coronal field lines are also present at the left-most part of the observed magnetoplasma system in the given field-of-view, however, their upper end is also diffused in the overlying corona (cf., Fig.~1, right-column, 09:20:09 UT). 

To mimic the physical scenario and dynamics of the different magnetic structures in the off-limb corona, we draw a schematic to emphasize on the
erupting prominence, overlying coronal magnetic fields, the onset of the forced magnetic reconnection, and associated plasma dynamics (Fig.~3). 
It describes observations vis-\'a-vis the dynamical evolution and overall configuration of the magnetic system containing a prominence, their activation, and dynamics of the overlying large-scale magnetic fields of the diffused corona (green and blue lines; top panel). The green lines depict the large-scale coronal magnetic fields lying just above the prominence system. The prominence associated with cool plasma just lies below this set of coronal magnetic field lines (i.e., green lines). The whole prominence material is bound by the low lying arcades (yellow arcs) crossing across its pillars/footpoints. One hot segment of the prominence (middle one among the orange colour prominence pillars) lifts up and triggers the expansion of the entire prominence system and subsequently expand and stretches the coronal magnetic field lines (top-right panel, green lines). It forcibly reconnects with the overlying diffused large-scale coronal magnetic field lines (blue lines). The onset of the forced reconnection is shown in the bottom-left panel, which is accompanied by the bi-directional multi-temperature plasma flows in the eastward and westward magnetic channels. The overlying regions above the reconnection point also open up into the overlying diffused corona, which enables the hot plasma flows in the upward direction also. The bottom-right panel mimics the onset of the post reconnection scenario, where the plasma flows in the eastward magnetic channel and further interacts and reconnects with another localized prominence segment, its overlying fields (loop/arcade), and open field lines to trigger the jet-like eruptions. The overlying magnetic field breaks and further disappeared after the eruption of the jets. The overall physical scenario depicted by this scehmatic, is clearly observed and described in greater details in Figs.1-2 and Figs.4-5.
\begin{figure*}
\begin{center}
\hspace{-0.5cm}
\includegraphics[scale=0.7,angle=0,width=18cm,height=18.0cm,keepaspectratio]{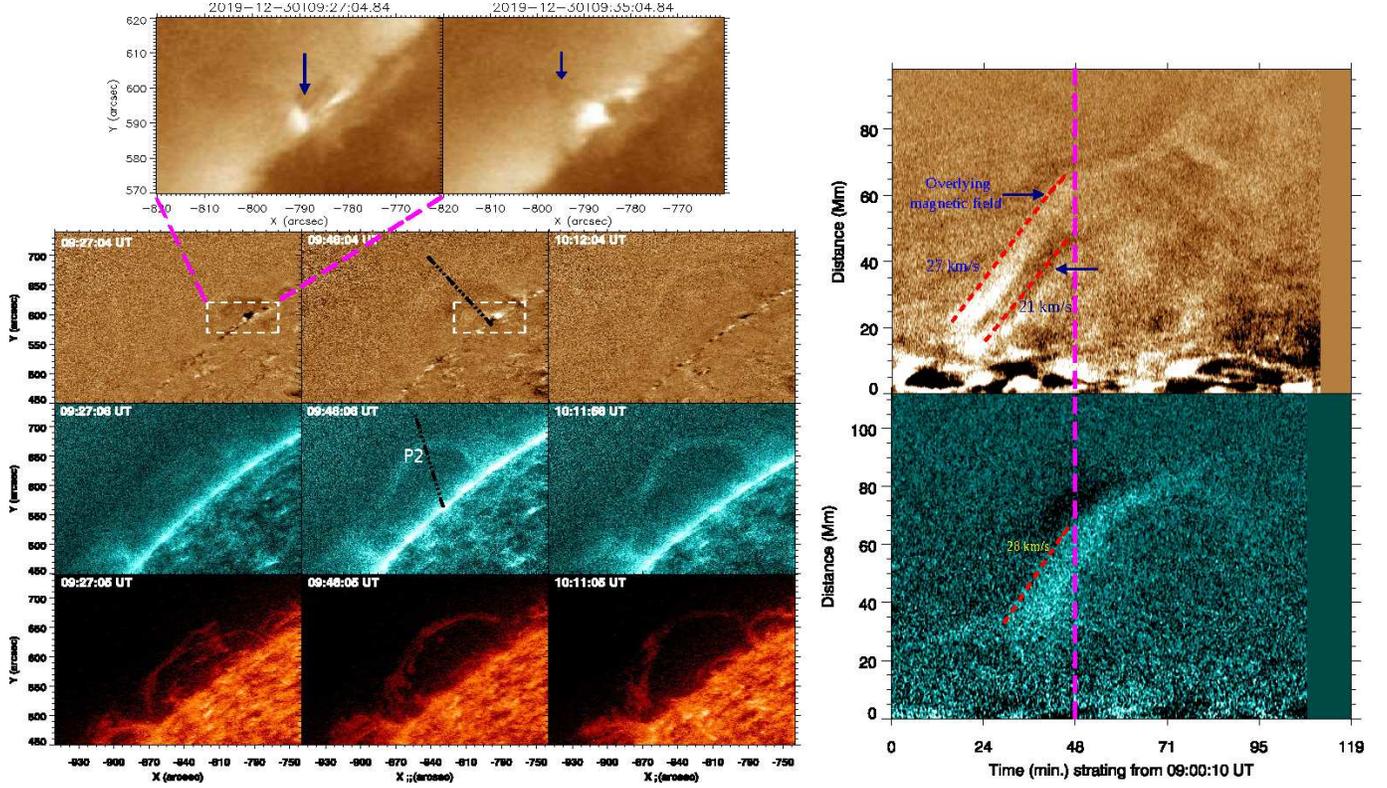}
\end{center}
\caption{Left panel: Multiwavelength imaging observations of SDO/AIA AIA 193 {\AA} (top), 131 {\AA} (middle), and 304 {\AA} (bottom) wavebands demonstrate the dynamical nature of the localized hot plasma segment of the prominence system, overlying magnetic fields embedded in the hot coronal plasma, below-lying cool prominence system, and overlying reconnection region in the large-scale off-limb corona. The top-left panels show that a hot plasma segment is erupted and trapped within the overlying magnetic domain. The AIA 193 {\AA} images show the expansion of the hot plasma segment and overlying magnetic field configuration. Top-right panel: A slit 'P1' has been taken on 193~\AA~ (top-left panel) image at 09:46 UT along the expansion of the localized hot plasma segment and overlying magnetic structures. The corressponding distance-time map along 'P1' estimates the kinematics of the hot plasma segment ($\approx$21 km s$^{-1}$) and overlying magnetic field ($\approx$27 km s$^{-1}$). Middle-left panel: The AIA 131 {\AA} determines the evolution of the same overlying magnetic field embedded in the hot coronal plasma just lying above the cool prominence. A slit 'P2' has been taken along the expansion of  these field lines to deduce velocity in 131~\AA~. Bottom-right panel: It is seen in the corressponding distance-time map in 131~\AA~ along 'P2' that the coronal field lines move with the speed of $\approx$28 km s$^{-1}$. This speed almost resembles with the same as observed in 193~\AA~waveband. Bottom-left panel: The AIA 304~\AA~images show the expansion and upward motion of the entire cool prominence system just below the overlying stretched coronal magnetic field configuration (top and middle-left panels).}
\end{figure*}
As explained above in the schematic, the magnetic complexity of the system is much larger in the present observational base-line (Figs. 1-2). The cool prominence system is embedded inside the overlying large-scale coronal magnetic field configurations (cf., Fig.~1, right-column, 09:20:09 and 09:46:09 UT). It is initially became unstable due to the localized eruption of the hot plasma segment of the prominence system (cf., Figs 1-2). This process further triggers the expansion of the entire cool prominence system that further stretches the overying coronal magnetic fields and enable them to be transported into the forced reconnection region in the off-limb corona. 
\begin{figure*}
\begin{center}
\hspace{-0.5cm}
\includegraphics[scale=0.7,angle=0,width=18cm,height=18.0cm,keepaspectratio]{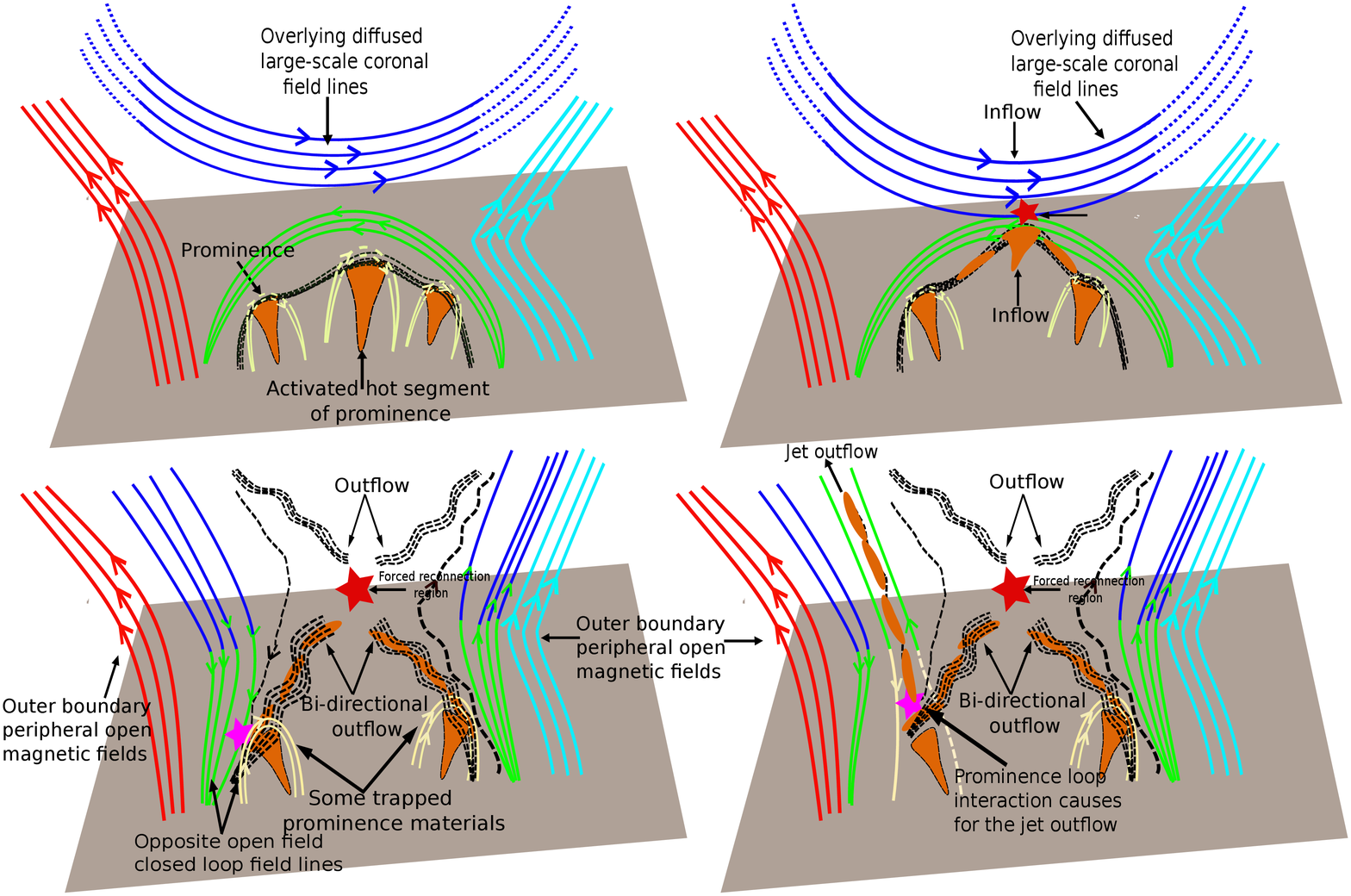}
\end{center}
\caption{A schematic displaying all the stages of the observed forced reconnection and associated plasma dynamics.}
\end{figure*}
\begin{figure*}
\begin{center}
\hspace{-2.0cm}
\includegraphics[scale=0.7,angle=0,width=15.5cm,height=16.0cm,keepaspectratio]{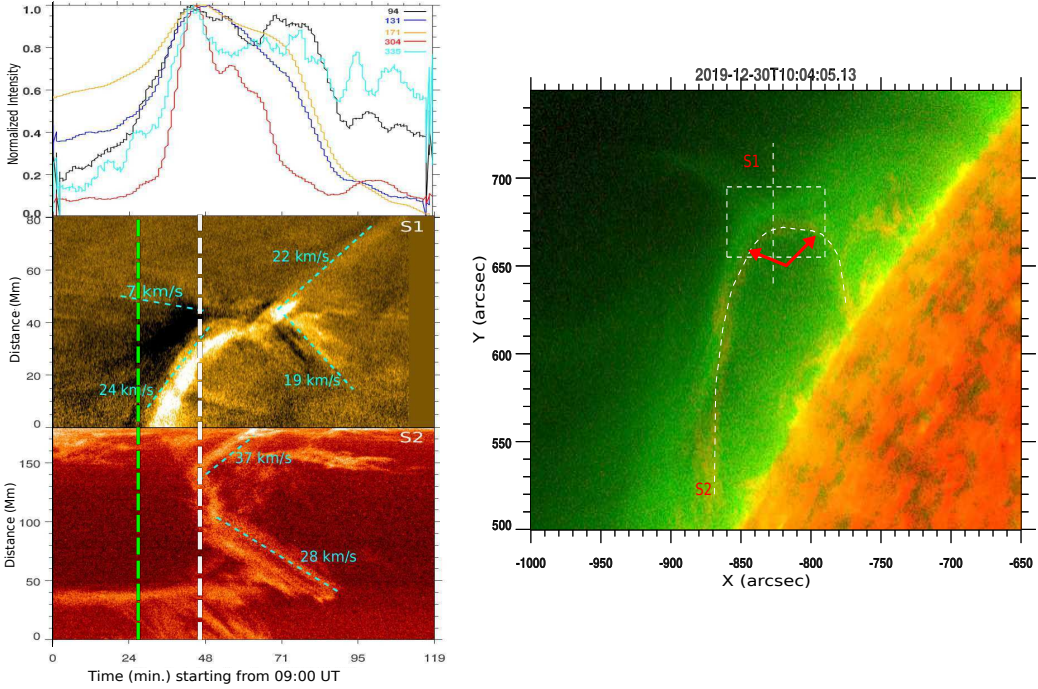}
\end{center}
\caption{Distance-time measurements: (i) along the path S1 (right-panel) that captures the inflowing plasma motions towards the reconnection point and some traces of upward plasma flows in the later phase of the reconnection which is triggered during the magnetic field reorganization, (ii) along a curved path S2 (right-panel) where the bi-direction plasma channelling takes place downward towards the limb during the forced reconnection. The (i) and (ii) are respectively displayed as distance-time maps in 171 {\AA} difference (middle-left panel) and 304 {\AA} (bottom-left panel). On top of it, the temporal variation of the emissions from the reconnection region (box shown in the right-panel) recorded in various AIA channels, have been plotted (top-left panel). The normalized intensities I=(I$_{o}$-I$_{max}$)/(I$_{max}$+I$_{min}$) are derived from 1.0 min cadence AIA data, which are also smoothed by '10' window-width to observe the long term variations. The green vertical line indicates initiation time of the inflow along slit 'S1'. The white vertical line indicates the onset timing of the forced magnetic reconnection.}
\end{figure*}
\begin{figure*}
\begin{center}
\includegraphics[scale=0.6,angle=0,width=18.5cm,height=19.0cm,keepaspectratio]{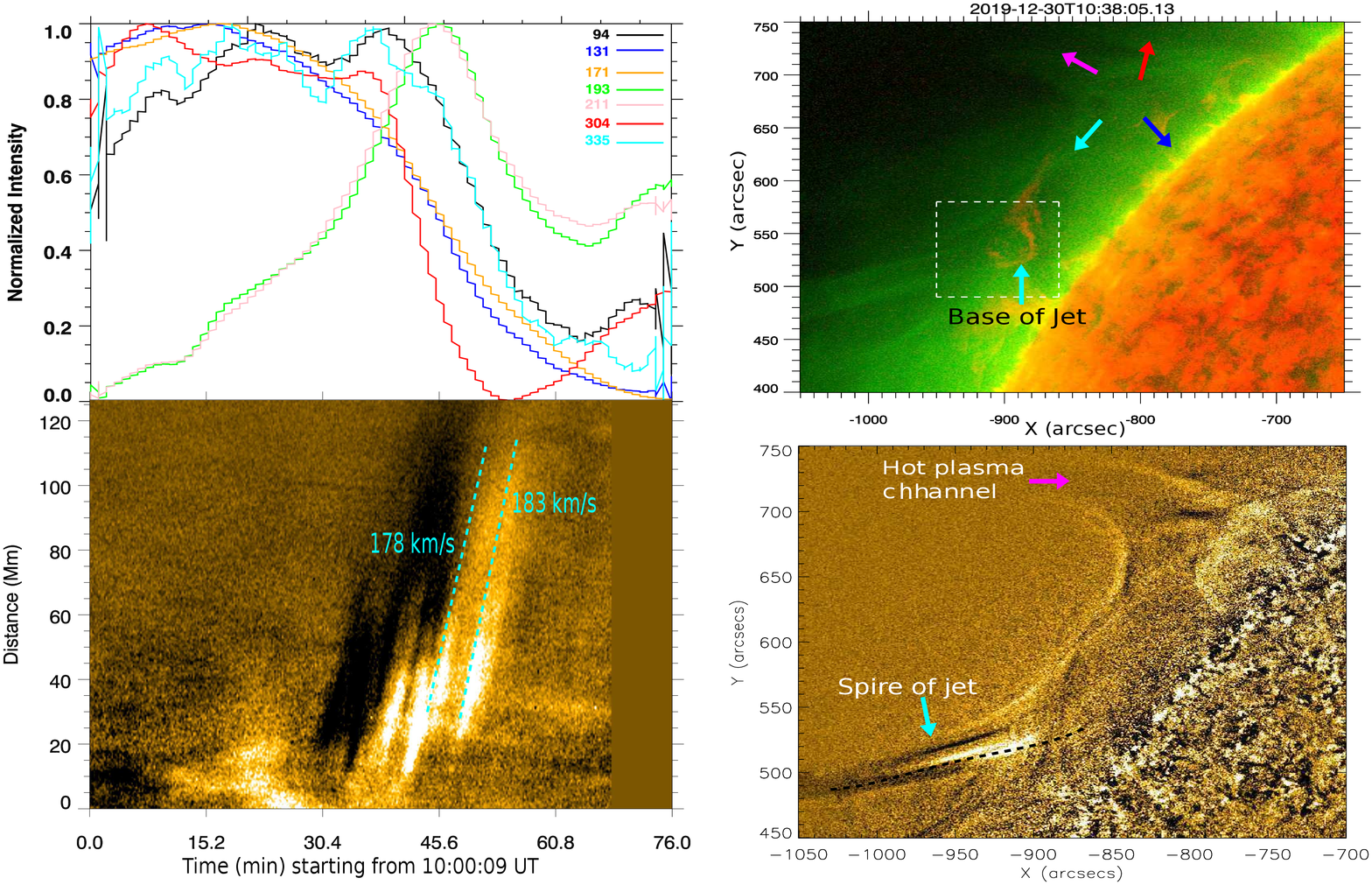}
\end{center}
\vspace{-0.6cm}
\caption{Top-right panel: Post-reconnection scenario at 10:38 UT when new eastward and westward magnetic configurations were already formed, and plasma outflows/channeling take place in upward (red and pink arrows) and downward (blue and cyan arrows) directions from the reconnection site. Bottom-right: The running difference image of 171 {\AA} also shows that new eastward and westward magnetic configurations were already formed in the post reconnection phase. At the left-most region, a jet-like spire and its structure is visible. Bottom-left: The distance-time diagram along the dashed black slit on the jet's spire shows that plasma moves impulsively multiple times constituting coronal jet-like eruptions. Top-left: The emissions as recorded in various AIA channels at the base region of the jet-like structure, which shows that these emissions peak multiple times whenever plasma is propelled up through the jet's spire.
The normalized intensities I=(I$_{o}$-I$_{max}$)/(I$_{max}$+I$_{min}$) are plotted using 1.0 min cadence AIA data, which are smoothed by '10' window-width to observe the long term variations.
The animation5d.mp4 (171) \AA~in the SDO/AIA running difference images shows the complete eruption of the prominence system, stretching of the overlying coronal magnetic field configurations in and around the reconnection region, associated plasma dynamics near the reconnection region, chanelling of the multi-temperature plasma in the post-reconnection phase, and triggering of the jets. It runs from 09:28 to 11:00 UT.}
\end{figure*}

Fig.~4 shows distance-time measurements (i) along the path S1 (right-panel), which estimates the kinematics of the inflowing plasma towards the reconnection point, and (ii) along a curved path S2 (right-panel) on which the bi-directional plasma channeling takes place during the forced reconnection once the magnetic re-organization takes place. These measurements are respectively displayed as distance-time maps in  171 {\AA} running difference and 304 {\AA} (left-panel). On top of it, the temporal variation of the emissions from the reconnection region (white dashed box in the right-panel) as recorded by various AIA channels, have been plotted. The distance-time map along S1 shows that plasma associated with the prominence system moves in the upward direction on 09:28 UT by the speed of 24 km s$^{-1}$. Almost on the same time less denser plasma moves inward from the top with an average speed of 7 km s$^{-1}$. This north-south motion of the field lines into a reconnection region occurs during 09:28 UT-09:48 UT. It should be noted that this is a projected scenario of the motion. Also the bi-directional north-south motion of the magnetoplasma system is significantly different in the speed here because the upward motion of the coronal plasma is forced by a prominence system from inward direction. This feature is itself different from the normal reconnection process, and a very typical of the forced magnetic reconnection \citep{2019ApJ...887..137S}. The reconnection takes place around $\approx$09:48 UT due to the forced plasma inflows as seen from north (top)- south (bottom) in two dimensional projection. Thereafter, the re-organization/re-orientation of the magnetic field also takes place in the East-West directions (cf., Fig.~1; animations 1, 2, \& 3).

After the forced reconnection, the right-most magnetic field domain is locally closed and right-most small prominence segment is trapped within it (Figs.~2\,--\,4). It is also seen that upper part of this magnetic domain is an open channel (cf., red and pink arrows in top-right panel of Fig.~5). In the difference image (bottom-right panel of Fig.~5) of 171 {\AA}, it is also clearly observed that well after the forced reconnection (10:18 UT-10:38 UT), the east-west magnetic domains are completely separated and newly configured right-most (westward) magnetic configuration/domain exhibits cusp like structure. While, left-most (eastward) magnetic domain is now constituted a separate plasma channel whose one end is going more eastward over the left-most prominence segment where later the jet-like eruption is formed (cf., Fig.~5, top-right \& bottom-right panels). Its upper end is curved and going up as an open magnetic channel in the overlying diffused corona (cf., pink arrows in Fig.~5, top-right \& bottom-right panels). These two magnetic channels open-up respectively as a separator between the eastward and westward newly configured magnetic field domains after the commencement of the forced reconnection, and re-orientation/re-organization of the magnetic fields took place.  

\begin{figure*}
\begin{center}
\includegraphics[scale=0.6,angle=0,width=18.5cm,height=19.0cm,keepaspectratio]{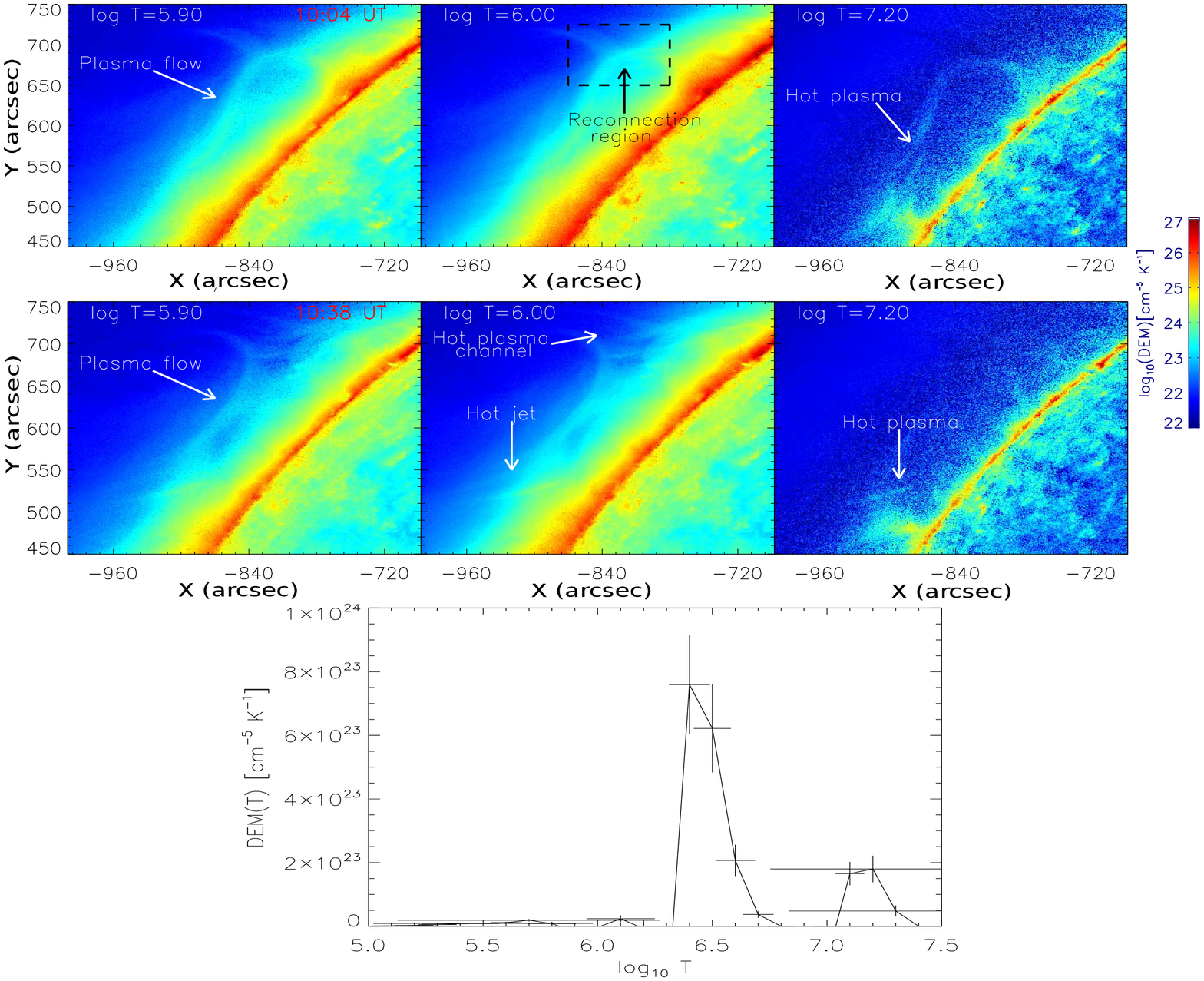}
\end{center}
\vspace{-0.3cm}
\caption{Top and Middle Panels: DEM maps at different temperatures of Log T$_{e}$= 5.9, 6.0, 7.2 at (i) 10:04 UT when the forced reconnection is gradually progressing, (ii) 10:38 UT when the responses in post-reconnection phase have been observed in form of multi-temperature plasma flows and onset of a jet-like eruption. Bottom Panel: The DEM vs temperature plot from the forced reconnection region.}
\end{figure*}

The analysis of the distance-time map created along the curved path 'S2' (Fig.~4, left-bottom panel) depicts that as soon as the inflow of two opposite magnetoplasma threads from north-south undergo into the forced reconnection at $\approx$9:48 UT, the east-westward re-orientation of the magnetic fields starts creating a bi-directional channeling of the plasma at 09:53 UT which is traced by its cool components (shown by red arrows in right-panel of Fig.~4). We have drawn two vertical lines (green and white) to indicate the initiation time of the inflow and outflow along slits 'S1' and 'S2' respectively. The northward-southward inflow starts when the heated plasma segment of the prominence system started to lift up (09:24 UT) and destablises and expands the entire cool prominence system towards the X-point from below regions. Once these dynamics are developed along north-southward direction, the inflow of the plasma also starts (09:28 UT) along slit 'S1' in the North-South direction. The forced reconnection begins at $\approx$09:48 UT (Fig.~4, white vertical line). After the onset of the forced reconnection, the plasma outflow is initiated along the curved path in East-West direction along the slit 'S2'. Their respective outflow speeds in right (westward) and left (eastward) directions are 37 km s$^{-1}$ and 28 km s$^{-1}$ respectively. The time lag of $\approx$20 min between the initiation of inflows and outflows is the characteristic observable of the forced reconnection \citep{2019ApJ...887..137S}. The estimated outflow speeds are the projected minimum speed, and their values depend upon the local magnetic field strength and plasma density into the reconnection region. Since, the forced reconnection region involves the bulky prominence and overlying coronal magnetic fields, therefore, due to its larger density, the outflow speeds must be significantly lower than the characteristic Alfv\'en speed in the inner corona \citep{2012A&A...540L..10I, 2019ApJ...874...57M,2019ApJ...887..137S}

As mentioned above that two magnetoplasma channels reconfigured/reorient in both the eastward and westward directions after the reconnection whose one end open up in the overlying diffused corona.  Therefore, along the slit 'S1' there are the traces (projected along the slit 'S1') of upward outflows of the coronal plasma with the speed of 22 km s$^{-1}$ and 19 km s$^{-1}$ that are also seen (cf., Fig.~4, left-middle panel). This complex scenario of the plasma motion is observed during the forced reconnection process.
The normalized intensities of various AIA filters as captured from the reconnection region are plotted in Fig.~4, left-top panel, which clearly describes that different emissions even sensitive to the high temperature plasma (e.g., 131, 094, 335 {\AA}) are peaked during the time of the forced reconnection at 09:48 UT. After the gradual progress of the reconnection, these emissions subside except in few AIA channels (e.g., 335, 094, and 171 {\AA}) because some traces of the hot plasma also flow along the upward open magnetic channels well after the reconnection.

After the commencement of the forced reconnection, the DEM map at Log T$_{e}$=7.2 on 10.04 UT (cf., Fig.~6 top-right panel) shows the existence of hot plasma also along the eastward and westward magnetic channels, propelling down towards the limb. It should be noted that multi-temperature plasma is created, and prominence material is also heated upto some extent to the transition region-inner coronal (T$_{e}$=5.9-6.0) temperature. The evolution of the temperature at the reconnection site is shown in the bottom-panel of Fig.~6. It is evident that most of the emissions from the reconnection region come from the plasma at approximately Log T$_{e}$=6.4, while its quarter fraction is coming from the plasma at Log T$_{e}$=7.2. The hot plasma mixed with the cool traces of prominence as well as plasma maintained at typical coronal temperature flow towards the eastward magnetic channel (cf., Fig.~6 top panels, and Fig.~5).

Fig.~5 (right-panel) and DEM maps in Fig.~6 (all middle panels) describe that after the reconnection the plasma outflows took place in the eastward direction towards left-most localized prominence system as seen in the field-of-view (Figs.~1\,--\,5; animations 1, 2, \& 3). The multi-temperature plasma, which consists of some flowing prominence threads and their magnetic fields too, hurls towards this prominence and compress this region. Reconnection between the prominence and overlying magnetic fields takes place and the base and spire of a coronal jet-like structure builds-up (cf., Fig.~5, top-right \& bottom-right panels). The distance-time diagram in Fig.~5 as well animations (animations 1, 2, \& 3) clearly show that the base of jet-like structure is evolved during 10:24 UT-10:38 UT by prominence-prominence interaction and prominence-loop (overlying arcade field) interaction \citep{2010ApJ...710.1195K,2016NatPh..12..847L}. Subsequent reconnection took place, and thereafter hot plasma is ejected multiple times along the spire or open magnetic channel (black-dashed line in the right-bottom panel of Fig.~5) with the high speed ranging from 178 to 183 km s$^{-1}$ (Fig.~5, bottom-left panel). DEM maps (Fig.~6, middle panels) also show the formation of hot jet-like spire at the left-most region through which the multi-temperature plasma (Log T$_{e}$=6.0-7.2) is propelled into the higher atmosphere. This jet-like propulsion is found to be highly impulsive with the triggering of high-speed plasma multiple times. The emissions in various AIA channels at the base region/lower segment of this jet-like structure show that they peak multiple times whenever plasma propelled through the jet's spire (Fig.~5, top-left panel). This is the indicative that multiple episodic reconnection are occurred at the base of this jet-like structure during prominence-prominence/loop interaction when multi-temperature plasma as well as the traces of the prominence were propelled from the forced reconnection region to this particular region. After the formation of jet, the overlying magnetic channel is disappeared in the overlying corona (cf., animations 1, 2, \& 3; Schematic in Fig.~3).
\vspace{-0.1cm}
\section{Discussion and Conclusions}

In this paper, we investigate the observational signature of the prominence driven forced reconnection in the off-limb large-scale corona. The eruptive prominence acts as an external driver to trigger the expansion and stretching of the overlying coronal magnetic field lines that undergo into the forced magnetic reconnection with the overlying another set of coronal magnetic field lines inflowing towards the reconnection region from the top in the off-limb corona. The direct observational manifestation of the forced reconnection, externally driven by a prominence, is given by \citet{2019ApJ...887..137S}. However, they did not observe a large temperature evolution at the site of the forced reconnection due to the fact that partially ionized, dense and collision dominated prominence plasma might consume the liberated energy \citep{2006ApJ...641.1217C}. Another point was that they have observed the non-flaring quiescent loop system far off-limb in the structured corona, where magnetic complexities were not present much. However, in the present case, the magnetic complexity was more predominant, and a cool prominence system and, stretched and expanding coronal magnetic field lines rose up near the limb to drive the forced reconnection with the overlying loop system. Moreover, in 2-D projection, the inflows appear to be squeezing~into the~reconnection region from north-south direction, and after the reconnection the new eastward-westward magnetic configuration/channel is generated and separated. This process has created a complex pattern of the outflowing plasma both in the upward direction along the open magnetic channel above the reconnection region, as well as in the downward direction respectively towards the footpoints of eastward and westward separated magnetic channels. Later, the multi-temperature plasma carries prominence threads also along with it towards the footpoint of the eastward magnetic channel, which further causes the prominence-prominence interaction and prominence-loop interaction to generate a coronal jet-like structure \citep{2010ApJ...710.1195K,2016NatPh..12..847L}. 

It is interesting to note that hot plasma flows appear through the eastward magnetic channel over a length of $\approx$10$^{5}$ km after the reconnection (cf., Fig.~6). The erupted prominence material is also channeled through the same set of the magnetic field lines. It is seen in the present observational base line that the embedded core prominence material and outer peripheral coronal magnetic fields/loop system directly interact with the overlying inflowing typical loop system within the reconnection region. The streching and expansion of the peripheral coronal magnetic field lines due to the eruption of the prominence from below further triggers the inflows and subsequent forced magnetic reconnection. However, the outflowing plasma consists of the mixture of multi-temperature hot plasma and remaining bulky prominence threads/materials. Prominence material did not directly absorb the generated energy at the reconnection site, and the coronal plasma is directly heated to the range of the temperature (i.e, Log T$_{e}= 6.0 - 7.2$) at the reconnection site. This physical effect is well demonstrated collectively in form of the elevation of normalized intensities of various AIA channels as shown in Fig.~4, DEM maps in the top-panel of Fig.~6, and DEM vs temperature plot in the bottom panel of Fig.~6. 

\citet{2014ApJ...788...85V} have observed the magnetic reconnection by an eruption of a filament and the expansion of its magnetic field in the corona. Similarly, \citet{1995JGR...100.3355F}, \citet{2000ApJ...545..524C}, and \citet{2009ApJ...697..913O} have observed that an emerging flux reconnects with the overlying coronal magnetic field and responsible for the large-scale eruptions. These above-mentioned observations belong to the category of spontaneous reconnection. In such findings, the expansion of the eruptive field lines associated with the flux emergence directly reconnect themselves with the overlying coronal magnetic fields to lead the magnetic reconnection. However, in the forced reconnection scenario, an additional step is basically evolved, which is seen also in the present observational study. An external perturbations (e.g., eruption, waves, release of photospheric magnetic field shearing, etc) may disturb or force the surrounding field lines to reconnect with the overlying or nearby oppositely directed magnetic fields \citep{2019ApJ...887..137S,2020A&A...643A.140M}. The dynamical nature of the Sun may support such externally driven reconnections more often and related physical scenario at different spatio-temporal scales \citep[e.g.,][and references cited there]{2016SSRv..200...75N,2017ApJ...847...98J,2019ApJ...887..137S,2020A&A...643A.140M}. There are few observational evidences of such a intriguing plasma processes in the solar atmosphere now \citep[e.g.,][]{2019ApJ...887..137S,2020A&A...643A.140M}. In the present paper, we provide a observational scenario of the forced reconnection which is occurred off the limb into the solar corona. It is driven by the eruption of a cool prominence system and the overlying/peripheral expanding \& stretched coronal field lines. These magnetic structures further jointly transported into the reconnection region and reconnect with the opposite coronal fields flowing inward from the top regions of the off-limb corona. The notable outcome of such a reconnection is the heating, as well as formation of hot and cool plasma motions both in the downward as well as upward directions from the reconnection site. Some secondary dynamical plasma processes, e.g., formation of the jet-like structure in the vicinity, are also seen at the later times in the same magnetic domain. Therefore, we establish the facts that such reconnection is forced and affect severely the localized coronal regions both in terms of enegetics as well as plasma dynamics. Such physical phenomena should be examined in details while we study the dynamical solar coronal/eruptive regions. 

As we are well aware that the concept of the forced reconnection is established in form of analytical theory and numerical modeling in variety of plasma, including solar and astrophysical plasma \citep[e.g.,][,references cited there]{1985PhFl...28.2412H,1998PhPl....5.1506V,2001PhPl....8..132B,2005GeoRL..32.6105B, 2005PhPl...12a2904J,2010RvMP...82..603Y,2017JPlPh..83e2001V,2017PhPl...24e2508B,2019A&A...623A..15P}. Recently, \citet{2019ApJ...887..137S} have firstly introduced that even in the presence of less amount of resistivity, the efficient forcing from the external driver implemented in the vicinity of an X-point, can trigger the reconnection at a reasonable rate. They have also observed the similar phenomenon in the large-scale solar corona directly. However, more stringent modeling of the complex magnetic field configuration mimicking the observed conditions (e.g., as seen in the present observations), role of the external drivers associated with the realistic eruptive conditions (e.g., a cool prominence eruption in the present case), and quantifying the heating and dynamical motions, are the frontline questions that need to be addressed ({cf., Figs.~1\,--\,5}). After the claim of the first direct observational signature of the forced reconnection in the large-scale solar corona by \citet{2019ApJ...887..137S}, this is the another remarkable case that depicts about some more complex magnetoplasma environment in the localized solar corona where forced reconnection took place by the eruption of a cool prominence system, and significant temperature evolution and plasma dynamics have been occurred in the solar corona. In conclusion, these meaningful multiwavelength observations provide the signature of a prominence driven forced reconnection, and demonstrate its role in heating the corona locally, and also in generating the hot plasma flows and jet-like motions. However, more observations and refined data-driven modeling of the forced reconnection should be performed, and its potential role in the heating and dynamics of the solar corona must be explored and established in the light of such novel observational findings.

\section{Acknowledgment}
We thank referee for his/her valuable comments that improved the manuscript. Authors acknolwedge UKIERI project grant, IIT (BHU) and SERB-DST financial support, and use of SDO/AIA observational data.
PJ acknowledges support from grant 21-16508J of the Grant Agency of the Czech Republic.


\end{document}